# Extremely strong-coupling superconductivity and anomalous lattice properties in the *β*-pyrochlore oxide KOs$_2$O$_6$

Z. Hiroi, S. Yonezawa, Y. Nagao and J. Yamaura

*Institute for Solid State Physics, University of Tokyo, Kashiwa, Chiba 277-8581, Japan*

Superconducting and normal-state properties of the β-pyrochlore oxide KOs$_2$O$_6$ are studied by means of thermodynamic and transport measurements. It is shown that the superconductivity is of conventional *s*-wave type and lies in the extremely strong-coupling regime. Specific heat and resistivity measurements reveal that there are characteristic low-energy phonons that give rise to unusual scattering of carriers due to strong electron-phonon interactions. The entity of the low-energy phonons is ascribed to the heavy rattling of the K ion confined in an oversized cage made of OsO$_6$ octahedra. It is suggested that this electron-rattler coupling mediates the Cooper pairing, resulting in the extremely strong-coupling superconductivity.

## I. INTRODUCTION

The study of non-Cu-based oxide superconductors has been extended during the last decade, aiming at understanding the role of electron correlations in the mechanism of superconductivity or searching for a novel pairing mechanism, hopefully to reach a higher $T_c$. An interesting example recently found is a family of pyrochlore oxide superconductors. The first discovered is α-pyrochlore rhenate Cd$_2$Re$_2$O$_7$ with $T_c$ = 1.0 K[1-3] and the second β-pyrochlore osmate $A$Os$_2$O$_6$ with $T_c$ = 3.3, 6.3, and 9.6 K for $A$ = Cs,[4] Rb,[5-7] and K,[8] respectively. They crystallize in the cubic pyrochlore structure of space group $Fd$-3$m$ and commonly possess a 3D skeleton made of ReO$_6$ or OsO$_6$ octahedra.[9] The "pyrochlore" sublattice occupied by the transition metal ions is comprised of corner-sharing tetrahedra that are known to be highly frustrating for a localized spin system with antiferromagnetic nearest-neighbor interactions.

A unique structural feature for the β-pyrochlores is that a relatively small $A$ ion is located in an oversized atomic cage made of OsO$_6$ octahedra, Fig. 1. Due to this large size mismatch, the $A$ atom can rattle in the cage.[10] The rattling has been recognized recently as an interesting phenomenon for a class of compounds like filled skutterudites[11] and Ge/Si clathrates[12] and attracted many researchers, because it may suppress thermal conductivity leading to an enhanced thermoelectric efficiency. On the other hand, the rattling is also intriguing from the viewpoint of lattice dynamics: it gives an almost localized mode even in a crystalline material and often exhibits unusual anharmonicity.[11] Hence, it is considered that the rattling is a new type of low-lying excitations that potentially affects various properties in a crystal at low temperature.

In the β pyrochlores, specific heat experiments found low-energy contributions that could be described approximately by the Einstein model and determined the Einstein temperature $T_E$ to be 70 K, 60 K, and 40 or 31 K for $A$ = Cs, Rb, and K, respectively.[13, 14] This tendency illustrates uniqueness of the rattling, because, to the contrary, one expects a higher frequency for a lighter atom in the case of conventional phonons. Moreover, it was demonstrated that the specific heat shows an unusual $T^5$ dependence at low temperature below 7 K for $A$ = Cs and Rb, instead of a usual $T^3$ dependence from a Debye-type phonon.[13] On one hand, structural refinements revealed large atomic displacement parameters at room temperature of $100U_{iso}$ = 2.48, 4.26, and 7.35 Å$^2$ for $A$ = Cs, Rb, and K, respectively,[10] and 3.41 Å$^2$ for Rb.[15] Particularly, the value for K is enormous and may be the largest among rattlers so far known in related compounds. This trend over the β-pyrochlore series is ascribed to the fact that an available space for the $A$ ion to move in a rather rigid cage increases with decreasing the ionic radius of the $A$ ion.[10] Kuneš *et al.* calculated an energy potential for each $A$ ion and found in fact a large anharmonicity, that is, a deviation from a quadratic form expected for the harmonic oscillator approximation.[16] Especially for the smallest K ion, they found 4 shallow potential minima locating away from the center (8$b$ site) along the <111> directions pointing to the nearest K ions, as schematically shown in Fig. 1. The potential minima are so shallow that the K ion may not stop at one of them even at very low temperature.

The electronic structures of α-Cd$_2$Re$_2$O$_7$ and β-$A$Os$_2$O$_6$ have been calculated by first-principle density-functional methods, which reveal that a metallic conduction occurs in the (Re, Os)-O network:[16-20] electronic states near the Fermi level originate from transition metal 5$d$ and O 2$p$ orbitals. Although the overall shape of the density of states (DOS) is similar for the two pyrochlores, a difference in band filling may result in different properties; Re$^{5+}$ for α-Cd$_2$Re$_2$O$_7$ has two 5$d$ electrons, while Os$^{5.5+}$ for β-$A$Os$_2$O$_6$ has two and a half. Moreover, a related α-pyrochlore Cd$_2$Os$_2$O$_7$ with Os$^{5+}$ (5$d^3$) exhibits a metal-to-insulator transition at 230 K.[21, 22]

Various experiments have been carried out on the pyrochlore oxide superconductors to elucidate the mechanism of the superconductivity. Most of the results obtained for α-Cd$_2$Re$_2$O$_7$ indicate that it is a weak-coupling BCS-type superconductor.[23, 24] In contrast, results on the β-pyrochlores are somewhat controversial. Although the $T_c$ increases smoothly from Cs to K, the jump in specific heat at $T_c$, the upper critical field, and the Sommerfeld coefficient all exhibit a large enhancement toward K.[25] Thus, the K compound is always distinguished from the others. Pressure dependence of $T_c$ was also studied for the two pyrochlores, showing a common feature: as pressure increases, $T_c$ first increases, exhibits a broad maximum and goes to zero above a critical pressure that depends on the system, for example, about 6 GPa for the K compound.[26-29]

On the symmetry of the superconducting gap for the β-pyrochlores, Rb-NMR and μSR experiments gave evidence for *s*-wave superconductivity for RbOs$_2$O$_6$.[30-32] In contrast, Arai *et al.* carried out K-NMR experiments and found no coherence peak in the relaxation rate below $T_c$ for KOs$_2$O$_6$, which seemed to indicate unconventional superconductivity.[32] However, their recent interpretation is that the absence of a coherence peak does not necessarily mean non-*s* pairing, because the relaxation rate probed by the K nuclei can be affected dominantly by strongly overdamped phonons.[33] Kasahara *et al.* measured thermal conductivity using a KOs$_2$O$_6$ single crystal and concluded a full gap from the insensitivity of thermal conductivity to magnetic fields.[34] Moreover, very recent photoemission spectroscopy (PES) experiments revealed the opening of a large isotropic gap below $T_c$.[35] On one hand, a μSR experiment claimed that the gap of $KOs_2O_6$ is anisotropic, or otherwise, there are two gaps.[36] Therefore, the pairing symmetry of the β-$AOs_2O_6$ superconductors may be of the conventional s wave, aside from minor aspects such as anisotropy or multi gaps, which means that the fundamental pairing mechanism is ascribed to phonons. Then, an important question is what kind of phonons are relevant for the occurrence. To find out the reason of the observed singular appearance toward K in the series must be the key to understand interesting physics involved in this system.

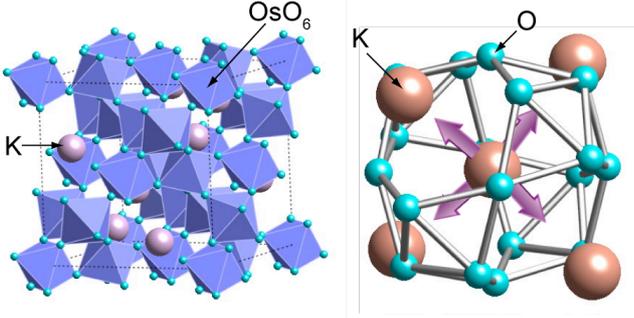

FIG. 1. (Color online) Crystal structure of the β-pyrochlore oxide $KOs_2O_6$. The K ion (big ball) is located in an oversized atomic cage made of $OsO_6$ octahedra and can move along the 4 <111> directions pointing to the neighboring K ions in adjacent cages.

Previous studies have suffered poor quality of samples, because only polycrystalline samples were available. The Sommerfeld coefficient $\gamma$ was estimated from specific heat by extracting contributions from Os metal impurity to be 40 mJ $K^{-2}$ $mol^{-1}$ for both Cs and Rb,[13] and 34 mJ $K^{-2}$ $mol^{-1}$ for Rb.[7] Recently, Brühwiler et al. obtained large values of $\gamma$ = 76-110 mJ $K^{-2}$ $mol^{-1}$ for $KOs_2O_6$ by collecting five dozen of tiny crystals.[14] However, there is an ambiguity in their values, because of uncertainty in their extrapolation method. They also reported strong-coupling superconductivity with a coupling constant $\lambda_{ep}$ = 1.0-1.6.[14] Recently, we successfully prepared a large single crystal of 1 mm size for $KOs_2O_6$ and reported two intriguing phenomena: one is a sharp and huge peak in specific heat at $T_p$ = 7.5-7.6 K below $T_c$, indicative of a first-order structural transition,[37,38] and the other is anisotropic flux pinning at low magnetic fields around 2 T.[39] It was suggested that the former is associated with the rattling freedom of the K ion. Moreover, anomalous concave-downward resistivity was observed down to $T_c$, suggesting a peculiar scattering mechanism of carriers.

In this paper, we present specific heat, magnetization and resistivity measurements on the same high-quality single crystal of $KOs_2O_6$. Reliable data on the superconducting and normal-state properties are obtained, which provide evidence for an extremely strong-coupling superconductivity realized in this compound. We discuss the role of rattling vibrations of the K ion on the mechanism of the superconductivity.

## II. EXPERIMENTAL

### A. Sample preparation

A high-quality single crystal was prepared and used for all the measurements in the present study, which was named KOs-729 after the date of July 29, 2005 when the first experiment was performed on this crystal. It was grown from a pellet containing an equimolar mixture of $KOsO_4$ and Os metal in a sealed quartz tube at 723 K for 24 h. Additional oxygen was supplied by using the thermal decomposition of AgO placed away from the pellet in the tube. The $KOsO_4$ powder had been prepared in advance from $KO_2$ and Os metal in the presence of excess oxygen. It was necessary to pay attention to avoid the formation of $OsO_4$ in the course of preparation, which is volatile even at room temperature and highly toxic to eyes or nose. After the reaction, several tiny crystals had grown on the surface of the pellet. Although the mechanism of the crystal growth has not yet been understood clearly, probably it occurs through partial melting and the following reaction with a vapor phase.

The KOs-729 crystal possesses a truncated octahedral shape with a large (111) facet as shown in Fig. 2 and is approximately 1.0 × 0.7 × 0.3 $mm^3$ in size and 1.302 mg in weight. The high quality of the crystal has been demonstrated by a sharp peak at $T_p$ in specific heat,[38] which was absent in the previous polycrystalline samples or appeared as broad humps in our previous aggregate of tiny crystals[37] or in five dozen of tiny crystals by Brühwiler et al.[14] Moreover, a dramatic angle dependence of flux-flow resistance was observed on this crystal, indicating that flux pinning is enhanced in magnetic fields along certain crystallographic directions such as [110], [001], and [112].[39] This evidences the absence of domains in this relatively large crystal.

A special care has been taken to keep the crystal always in a dry atmosphere, because it readily undergoes hydration in air, as reported in isostructural compounds such as $KNbWO_6$.[40,41] Once partial hydration takes place in $KOs_2O_6$, the second sharp anomaly in specific heat tends to collapse. In contrast, the superconducting transition was robust, just slightly broadened after hydration. The hydration must be relatively slow in a single crystal compared with the case of polycrystalline samples, which may be the reason for the broad anomaly or the absence in previous samples.

### B. Physical-property measurements

Both specific heat and electrical resistivity were measured in a temperature range between 300 K and 0.4 K and in magnetic fields of up to 14 T in a Quantum Design Physical Property Measurement System (PPMS) equipped with a $^3$He refrigerator. The magnetic fields had been calibrated by measuring the magnetization of a standard Pd specimen and also by measuring the voltage of a Hall device (F.W. BELL, BHA-921). Specific heat measurements were performed by the heat-relaxation method. The KOs-729 crystal was attached to an alumina platform by a small amount of Apiezon N grease. In each measurement, heat capacity was obtained by fitting a heat relaxation curve recorded after a heat pulse giving a temperature rise of approximately 2%. The heat capacity of an addendum had been measured in a separate run without a sample, and was subtracted from the data. The measurements were done three times at each temperature with a scatter less than 0.3% at most.

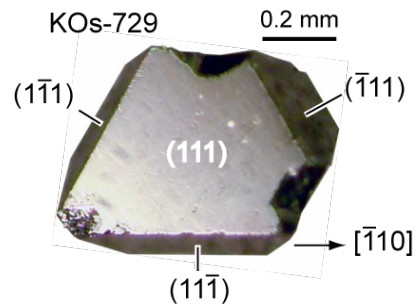

FIG. 2. (Color online) Photograph of the KOs-729 crystal used in the present study. It possesses a truncated octahedral shape with 111 facets. The approximate size is 1.0 × 0.7 × 0.3 $mm^3$. Resistivity measurements were carried out with a current flow along the [-110] direction.

Resistivity measurements were carried out by the four-probe method with a current flow along the [-110] direction and magnetic fields along the [111], [110], [001] or [112] direction of the cubic crystal structure. All the measurements were done at a current density of 1.5 A cm$^{-2}$. Magnetization was measured in magnetic fields up to 7 T in a Quantum Design magnetic property measurement system and also up to 14 T in PPMS. The magnetic fields were applied approximately along the [111] or [-110] direction.

## III. RESULTS

### A. Superconducting properties

#### 1. Specific heat

First of all, we analyze specific heat data in order to obtain a reliable value of the Sommerfeld coefficient $\gamma$. There are two obstacles: one is the large upper critical field $H_{c2}$ that is approximately 2 times greater than our experimental limit of 14 T. Brühwiler et al. reported $\gamma$ = 76 (110) mJ K$^{-2}$ mol$^{-1}$ assuming $\mu_0 H_{c2}$ = 24 (35) T by an extrapolation method.[14] Their values should be modified to $\gamma \sim$ 100 mJ K$^{-2}$ mol$^{-1}$, because recent high magnetic field experiments revealed $\mu_0 H_{c2}$ = 30.6 T or 33 T.[42, 43] Nevertheless, there are still large ambiguity in their extrapolation method using specific heat data obtained only at $H/H_{c2} <$ 0.5. The other difficulty comes from unusual lattice contributions in specific heat at low temperature and the existence of a sharp peak at $T_p$. Thus, it is not easy to extract the lattice contribution in a standard way used so far. Here we carefully analyze specific heat data and reasonably divide them into electronic and lattice parts, from which a reliable value of $\gamma$ is determined, and information on the superconducting gap is attained.

Figure 3 shows the temperature dependence of specific heat of the KOs-729 crystal measured on cooling at zero field and in a magnetic field of 14 T applied along the [111] direction. A superconducting transition at zero field takes place with a large jump, followed by a huge peak due to the second phase transition at $T_p$ = 7.5 K. The entropy-conserving construction shown in the inset gives $T_c$ = 9.60 K, $\Delta C/T_c$ = 201.2 mJ K$^{-2}$ mol$^{-1}$, which is close to the values previously reported,[14, 37] and a transition width ($\Delta T_c$) of 0.3 K. $T_c$ is reduced to 5.2 K at 14 T, which is evident as a bump in the 14-T data shown in Fig. 3. In contrast, Brühwiler et al. reported $T_c$ = 6.2 K at 14 T, though the transition was not clearly observed in their specific heat data. The $C/T$ at zero field rapidly decreases to zero as $T$ approaches absolute zero. The absence of a residual $T$-linear contribution in specific heat indicates the high quality of the crystal.

The specific heat of a crystal ($C$) is the sum of an electronic contribution ($C_e$) and an $H$-independent lattice contribution ($C_l$). The former becomes $C_{en}$ for the normal state above $T_c$, which is taken as $\gamma T$, and $C_{es}$ for the superconducting state below $T_c$. The $\gamma$ is assumed to be $T$-independent, though it can not be the case for compounds with strong electron-phonon couplings.[44] In the case of KOs$_2$O$_6$, $C_l$ is large relative to $C_e$: for example, $C_l$ is as large as ~90% of the total $C$ at just above $T_c$, as shown later. In order to determine the value of $\gamma$, it is crucial to know the low-temperature form of $C_l$. Since the minimum $T_c$ attained at 14 T is 5.2 K, one has to estimate the $C_l$ from the $T$ dependence of the total $C$ above ~5.5 K. Two terms in the harmonic-lattice approximation are often required for an adequate fit; $C_l = \beta_3 T^3 + \beta_5 T^5$. The first term comes from a Debye-type acoustic phonon, and thus is dominant at low temperature, while the second term expresses a deviation at high temperature. Actually, this approximation is valid for $\alpha$-Cd$_2$Re$_2$O$_7$, where $\beta_3$ = 0.222 mJ K$^{-4}$ mol$^{-1}$ and $\beta_5$ = 2.70 × 10$^{-6}$ mJ K$^{-6}$ mol$^{-1}$ are obtained by a fit to the data below 10 K.[1] The Debye temperature $\Theta_D$ is 458 K from the $\beta_3$ value. In strong contrast, it was found for two members of $\beta$-$A$Os$_2$O$_6$ that the $T^5$ term prevails in a wide temperature range; $\beta_5$ = 14.2 × 10$^{-3}$ mJ K$^{-6}$ mol$^{-1}$ below 5 K for CsOs$_2$O$_6$ and $\beta_5$ = 30.2 × 10$^{-3}$ mJ K$^{-6}$ mol$^{-1}$ below 7 K for RbOs$_2$O$_6$.[13]

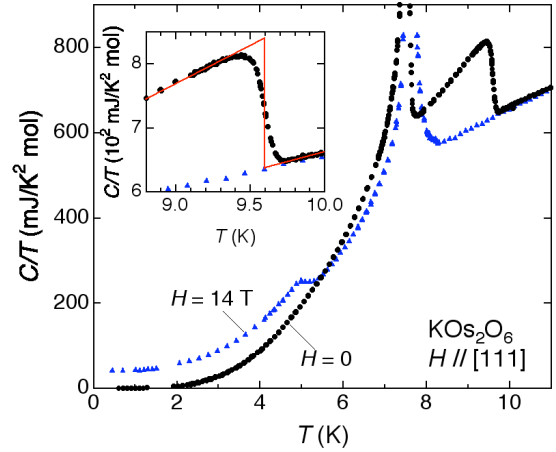

FIG. 3. (Color online) Specific heat divided by temperature measured at zero field (circle) and a magnetic field of 14 T (triangle) applied along the [111] direction. The inset shows an enlargement of the superconducting transition with an entropy-conserving construction.

The $C/T$ at $H$ = 0 below 7 K shown in Fig. 3 is again plotted in two ways as functions of $T^2$ and $T^4$ in Fig. 4. It is apparent from the $T^2$ plot that possible $T^3$ terms expected for $\Theta_D$ = 458 K and 300 K are negligibly small compared with the whole magnitude of specific heat, just as observed in other members. On the other hand, in the $T^4$ plot, there is distinct linear behavior at low temperature below 4 K, indicating that the $C$ approaches asymptotically to $T^5$ behavior as $T \rightarrow$ 0 with a large slope of 0.3481(6) mJ K$^{-6}$ mol$^{-1}$. It is reasonable to ascribe this $T^5$ contribution to the lattice, because $C_{es}$ should decrease quickly as $T \rightarrow$ 0. Note that the value of the $\beta_5$ for KOs$_2$O$_6$ is more than one order larger than those in other members. At high temperatures above 4 K, a downward deviation from the initial $T^5$ behavior is observed in Fig. 4b. The temperature dependence of the 14-T data above 5.5 K, which is taken as $\gamma T + C_l$, is also close to $T^5$, but with a smaller slope, which means that a single $T^5$ term is not appropriate to describe the $C_l$ in such a wide temperature range and also that an inclusion of higher order term of $T^n$ is not helpful. Therefore, we adopt expediently an alternative empirical form to express this strange lattice contribution; $C_l = \beta_5 T^5 f(T)$, where $f(T)$ = [1 + exp(1 - $pT^{-q}$)]$^{-1}$. Since the $f(T)$ is almost unity below a certain temperature and decreases gradually with increasing $T$, this $C_l$ can reproduce $T^5$ behavior at low temperatures and a weaker $T$ dependence at high temperatures. As shown in Fig. 4b, the 14-T data in the 5.5-7 K range can be fitted well by the function for a value of $\beta_5$ fixed to the initial slope of 0.3481 mJ K$^{-6}$ mol$^{-1}$ and a given value of $\gamma$, for example, $p = 6.96(3)$ and $q = 1.09(1)$ for $\gamma = 70$ mJ K$^{-2}$ mol$^{-1}$.

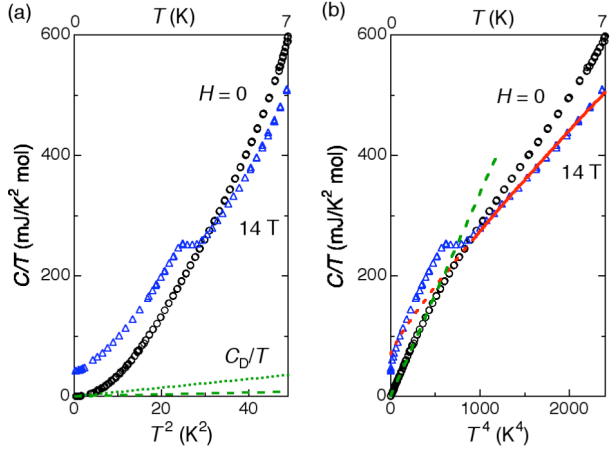

FIG. 4. (Color online) Low-temperature specific heat below 7 K plotted as functions of $T^2$ (a) and $T^4$ (b). The broken and dotted lines in (a) show calculated contributions from Debye $T^3$ phonons of $\Theta_D = 460$ K and 300 K, respectively, which are much smaller than the experimental values. The broken line in (b) is a linear fit to the zero-field data as $T \to 0$, which gives a coefficient of the $T^5$ term, $\beta_5 = 0.3481(6)$ mJ K$^{-6}$ mol$^{-1}$. The solid and dotted line is a fit to $C_l = \beta_5 T^5 f(T)$. See text for detail.

In order to determine the value of $\gamma$ univocally, the entropy conservation is taken into account for the 14-T data, as shown in Fig. 5: since the normal-state specific heat expected for the case of $T_c = 0$ is given by $(C_{en} + C_l)$ (dotted line in Fig. 5), the integration of $[C_{en} + C_l - C(14\ \text{T})]/T$ should become zero due to entropy balance. It is shown in the inset to Fig. 5 that the integrated value changes almost linearly with $\gamma$ and vanishes around $\gamma = 70$ mJ K$^{-2}$ mol$^{-1}$. Hence, one can determine the value of $\gamma$ unambiguously. A certain ambiguity may arise from the assumed lattice function. However, since the temperature dependence of $C_l$ is substantially weak in the $T$ range of interest, a possible correction on the $\gamma$ value must be minimal, say, less than 1 mJ K$^{-2}$ mol$^{-1}$.

Next we determine $C_{es}$ at zero field by subtracting the $C_l$ estimated above. The temperature dependence of $C_{es}$ does have the BCS form, $a\exp(-\Delta/k_B T_c)$, as shown in Fig. 6. The energy gap $\Delta$ obtained by fitting is 22.5 K, which corresponds to $2\Delta/k_B T_c = 4.69$, much larger than the BCS value of 3.53. The above $C_{es}$ at low temperature below 7 K is again plotted in Fig. 7 together with high-temperature $C_{es}$ above 5.5 K, which is obtained by subtracting the 14-T data from the zero field data as $C_{es} = C(0) - C(14\ \text{T})$. The two data sets obtained independently overlap well in the 5.5-7 K range, assuring the validity of the above analyses. Because of the existence of the second peak and its small shifts under magnetic fields, the data between 7 K and 8.3 K is to be excluded in the following discussion. Taking $\gamma = 70$ mJ K$^{-2}$ mol$^{-1}$, the jump in specific heat at $T_c$, $\Delta C/\gamma T_c$, reaches 2.87, much larger than 1.43 expected for a weak-coupling superconductor, indicating that KOs$_2$O$_6$ lies in the strong-coupling regime. Comparisons to other typical strong-coupling superconductors are made in section IV-B.

Here we analyze the data based on the $\alpha$ model that was developed to provide a semi-empirical approximation to the thermodynamic properties of strong-coupled superconductors in a wide range of coupling strengths with a single adjustable parameter,

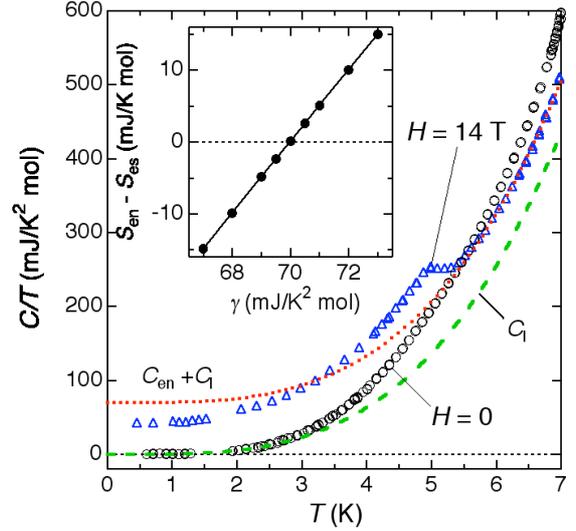

FIG. 5. (Color online) Specific heat data same as shown in Fig. 3. The dotted line shows the estimated contribution of $C_{en} + C_l$, and the broken line represents $C_l$ in the case of $\gamma = 70$ mJ K$^{-2}$ mol$^{-1}$. The inset shows a change of entropy balance as a function of $\gamma$, from which the value of $\gamma$ is decided to be 70 mJ K$^{-2}$ mol$^{-1}$.

$\alpha = \Delta_0/k_B T_c$.[45] Recently, it was generalized to a multi-gap superconductor and successfully applied to the analyses on MgB$_2$ or Nb$_3$Sn.[46,47] Using the $\alpha$ model, the data in the vicinity of $T_c$ is well reproduced, as shown in Fig. 7, and we obtain $\alpha = 2.50$ ($2\Delta_0/k_B T_c = 5.00$), slightly larger than the value obtained above from the temperature dependence of $C_{es}$. One interesting point to be noted is that there is a significant deviation between the data and the fitting curve at intermediate temperatures, suggesting the existence of an additional structure in the gap. Presumably, this enhancement would be explained if one assumes the coexistence of another smaller gap (not so small as in MgB$_2$, but intermediate). This possibility has been already pointed out in the previous $\mu$SR experiment.[36] However, ambiguity associated with the second peak in the present data prevents us from further analyzing the data. This important issue will be revisited in future work, where the

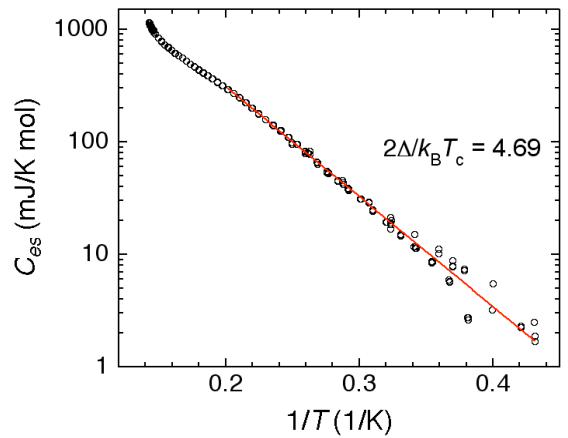

FIG. 6. (Color online) Temperature dependence of electronic specific heat measured at $H = 0$ for the superconducting state showing an exponential decrease at low temperature. A magnitude of the gap obtained is $2\Delta/k_B T_c = 4.69$.